\documentclass[pra,preprint,showkeys,showpacs,amsmath,amssymb]{revtex4}
\usepackage{bm}
\usepackage{graphics}

\newcommand{\vare}{\varepsilon }

\begin{document}

\title{Solitons in tunnel-coupled repulsive and attractive condensates}
\author{Valery S. Shchesnovich$^{1,2}$ }
\email{valery@ift.unesp.br}
\author{Solange B. Cavalcanti$^{2}$}
\email{solange@lux.ufal.br}
\author{Roberto A. Kraenkel$^1$}
\email{kraenkel@ift.unesp.br}

\affiliation{ $^1$Instituto de F\'{\i}sica Te\'{o}rica,
Universidade Estadual Paulista - UNESP,
Rua Pamplona 145, 01405-900 S\~{a}o Paulo, Brazil\\
$^2$Departamento de F\'{\i}sica - Universidade Federal de Alagoas,
Macei\'o AL 57072-970, Brazil }

\begin{abstract}

We study  solitons in the condensate trapped in a double-well potential
with far-separated wells, when the $s$-wave scattering length has different
signs in the two parts of the condensate. By employing the coupled-mode
approximation it is shown that there are unusual stable bright solitons in
the condensate, with the larger share of atoms being gathered in the
repulsive part. Such unusual solitons derive their stability from the
quantum tunneling and correspond to the strong coupling between the parts
of the condensate. The ground state of the system, however, corresponds to
weak coupling between the condensate parts, with the larger share of atoms
being gathered in the attractive part of the condensate.

\end{abstract}



\maketitle

\section{Introduction}
\label{sec1}

Bose-Einstein condensates (BECs) in trapped dilute gases exhibit interplay
between quantum and nonlinear phenomena, since at zero temperature the
mean-field Gross-Pitaevskii (GP) equation \cite{GPE} for the order
parameter applies with a good accuracy. The macroscopic quantum coherence
of BEC was demonstrated experimentally \cite{intrfexp,intrf2comp} and
explained theoretically \cite{intrfth} with the use of the GP equation (see
also the review Ref.~\cite{BECRev}). One of the manifestations of nonlinear
dynamics in BEC is appearance of solitons, i.e. the self-localized ``waves
of matter'', in quasi one-dimensional (cigar-shaped) condensates.

Nonlinear phenomena in BECs, and the solitons in particular, are similar to
that in nonlinear optics, for instance,  in optical fibers \cite{optfiber}.
In both  fields the nonlinear Schr\"odinger (NLS) equation appears at some
level of approximation.  Similar to optics, where bright and dark solitons
are supported respectively by the focusing and defocusing nonlinearities,
in BECs the $s$-wave scattering length is the determining factor. Dark
solitons are now routinely observed in the condensates with repulsive
interactions \cite{dark1,dark2,dark3,dark4}, while the condensates with
attractive interactions allow for stable bright solitons
\cite{bright1,bright2}.

The similarity with optics, however, does not go so far as the control over
nonlinearity in the governing equations.   Atomic interactions in BEC can
be tuned at one's will by application of magnetic field near the Feshbach
resonance \cite{fesh}. This opens a possibility to observe new phenomena in
BEC when the atoms interact differently in different spatial locations of
the trap due to the presence of an external field. Though such a setup has
not been yet realized experimentally, it cannot be rendered as impossible
at the theoretical level.  For BECs constrained to lower dimensions, the
Feshbach resonance proves to be sharp due to many-body effects, for
instance, it is so in the two-dimensional condensate \cite{Feshb2D}.
Similar result can be expected for the effectively one-dimensional BECs.
The feasibility of control over the scattering length in BEC by the optical
means  is also being discussed \cite{optcontr} (see also
Refs.~\cite{opt1,opt2}).

Control over the scattering length in a part of the condensate can be
realized in the double-well trap with far-separated wells, with the
gradient of the external field being localized in the region of the barrier
where the order parameter is exponentially small.  For such setup the
well-known two-mode approximation applies, which considerably simplifies
the governing equations and allows for the analytical study. The asymmetric
double-well potential is created by focusing an off-resonant intense laser
beam near the center of a parabolic magnetic trap. Such setup is routinely
realized in experiments (see, for instance, Ref.~\cite{intrfexp}).

The goal of the present paper is to understand  the nature of the unusual
bright solitons in the condensate trapped in an asymmetric double-well
potential, when the larger share of atoms is localized in the well where
the atomic interactions are vanishing or weakly repulsive. Such
counterintuitive solitons, due to the usual association of bright solitons
with attractive interactions, were recently numerically discovered
\cite{PhysD}. However, the analytical form of the solitons and the domain
of their existence were not addressed in the previous publication. In the
present paper we analytically study the solitons and  the domain of their
existence.

Nonlinear phenomena in BEC in the double-well trap have already attracted a
great deal of attention. The coherent atomic tunneling between the two
wells  and the macroscopic quantum self-trapping phenomenon were
theoretically predicted in Refs.~\cite{tunnel1,tunnel2,tunnellimit}. In
Ref.~\cite{tunnellimit}, the predictions of the mean field approach based
on the GP equation were also compared with the results from the full
quantum model.  The conclusion is that the characteristic time scale on
which the quantum collapse and revival dynamics starts to play a
significant role is large and increases with the number of BEC atoms, while
on the smaller time scale the GP equation gives a quantitatively accurate
description. Similar conclusions were derived recently in
Ref.~\cite{smalltunnl}. The general coupled-mode theory for the double-well
trap with strong tunneling was developed in Ref.~\cite{couplmod} using the
concepts of the nonlinear guided-wave optics.  Nonlinear modes of the
condensate in the double-well trap were used as the basis functions.
Finally, the possibility of observing a chaotic dynamics of the condensate
trapped in a double-well potential was the subject of Ref.~\cite{chaotdyn}.
An effective amplitude equation was derived, which exhibited a behavior
resembling that of the Lorentz system.

The paper is organized as follows. In the next section, complemented with
the appendix, we give a self-consistent derivation of the coupled mode
system for the asymmetric double-well trap with weak tunneling through the
barrier. The general properties of the solitons in the coupled-mode system
are summarized in  section \ref{genpr}. In section~\ref{sec3A}, we derive
an approximate analytical form of the  soliton solutions in the case of
vanishing interactions in one of the condensates. The exact boundary of the
soliton bifurcation form zero, which defines the domain of existence of the
stable solitons with the larger share of atoms in the repulsive condensate,
is found in section \ref{sec3B}. Section \ref{sec4}  contains discussion of
the predicted phenomena. Finally,  all numerical simulations in the present
paper were performed with the use of the spectral collocation methods
\cite{SP} known for their high-accuracy. More details can be found in
Ref.~\cite{PhysD}.

\section{Reduction of the GP equation to the coupled-mode system}
\label{sec2}

\begin{figure}[ht]
\includegraphics{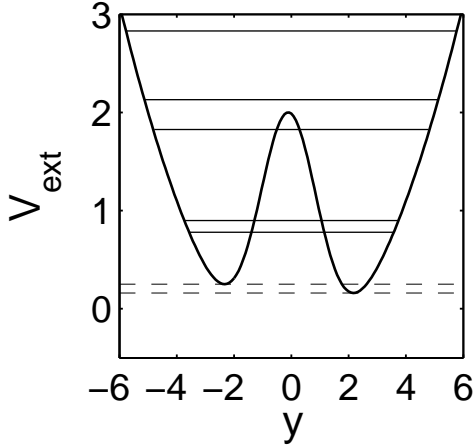}
\caption{\label{fg1} Illustration of the double-well: the parabolic trap
potential in the $y$-direction modified by a Gaussian barrier. The
(horizontal) solid lines indicate the numerically computed eigenvalues in
the combined potential. The dashed lines show the bottom energies in the
two wells of the trap. We use dimensionless units.}
\end{figure}

Consider BEC trapped in an asymmetric double-well potential and assume that
each of the wells is weakly confining in the $x$-direction and strongly
confining in the transverse plane $\vec{r}_\perp\equiv(y,z)$. Let the
transverse trap be parabolic in the $z$-direction and have the double-well
shape in the $y$-direction, see fig.~\ref{fg1}.

The zero-point energy difference between the two wells (shown by
the dashed lines in fig.~\ref{fg1}) can be managed, moving the
position of the barrier-generating laser beam.  We can neglect the
small variation of the trap potential in the $x$-direction. The
corresponding Gross-Pitaevskii (GP) equation for the order
parameter $\Psi$ of the condensate then reads
\begin{equation}
i\hbar \frac{\partial \Psi}{\partial t} =
\left\{-\frac{\hbar^2}{2m}\left(\frac{\partial^2 }{\partial x^2}+
\nabla_{\vec{r}{}_\perp}^2\right) +
\mathcal{V}_\mathrm{ext}(\vec{r}_\perp) + g|\Psi|^2\right\}\Psi,
\label{eq1}\end{equation}
 with the external potential
\begin{equation}
\mathcal{V}_\mathrm{ext} = \frac{m\omega_\perp^2}{2}(y^2 +
\gamma^2z^2)+\mathcal{V}_B\exp\left\{-\frac{(y-y_0)^2}{2\sigma^2}\right\},
\quad \mathcal{V}_B>0. \label{eq2}\end{equation}
 Here the parameter $\gamma$ accounts for the  trap asymmetry
in the transverse plane, while the parameters $y_0$ and $\sigma$ give the
position and the width of the barrier, respectively.

\begin{figure}[ht]
\includegraphics{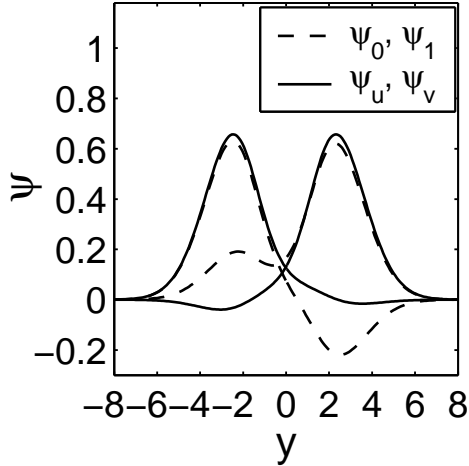}
\caption{\label{fg2} The ground and first excited state wave functions for
the double-well trap in fig.~\ref{fg1} (dashed lines) and the wave
functions localized in the complementary wells (solid lines) constructed as
indicated in the text. All units are dimensionless. }
\end{figure}

The double-well potential $\mathcal{V}_{\mathrm{ext}}$ is assumed to have
two quasi-degenerate energy levels:
\begin{equation}
E_1-E_0\ll E_2-E_1,
 \label{conda1}\end{equation}
fig.~\ref{fg1} can serve as an illustration. There is a natural choice of
the basis wave functions in the degenerate subspace, say $\psi_u$ and
$\psi_v$, where each one is localized in just one of the wells. Such basis
is given  by a rotation of the wave functions for the ground and the first
excited states:
\begin{equation}
\psi_u(\vec{r}_\perp) =  \frac{\psi_0(\vec{r}_\perp) +
\varkappa\psi_1(\vec{r}_\perp)}{\sqrt{1 + \varkappa^2}},\quad
\psi_v(\vec{r}_\perp) = \frac{\varkappa\psi_0(\vec{r}_\perp)-
\psi_1(\vec{r}_\perp)}{\sqrt{1 + \varkappa^2}}.
  \label{eq3}\end{equation}
 The choice of the parameter $\varkappa$ is based on the fact that
the quotients of the absolute values of the eigenfunctions
$\psi_0(\vec{r}_\perp)$ and $\psi_1(\vec{r}_\perp)$ at the extremal points
$y_-$ and $y_+$ (in the two wells) are in approximate inverse
proportionality relation
\begin{equation}
\frac{\psi_1(y_-)}{\psi_0(y_-)} \approx -\frac{\psi_0(y_+)}{\psi_1(y_+)}
\label{eq4}\end{equation}
 (the positions of the extremals of the wave functions
slightly deviate from each other and from the minima of the trap; these
deviations we neglect). Equation (\ref{eq4}) holds for the basis functions
of the degenerate subspace for a general asymmetric double-well trap with
weak tunneling through the barrier (see the appendix).  For the double-well
of fig.~\ref{fg1} the quotients of the eigenfunctions shown in
fig.~\ref{fg2} have the values $3.1$ at the left and $-2.9$ at the right
minima of the trap. We set $\varkappa = {\psi_1(y_-)}/{\psi_0(y_-)}$.
 For this choice of $\varkappa$ the wave functions $\psi_u(y)$ and $\psi_v(y)$
defined by equation (\ref{eq3}) are localized in the left and right wells,
respectively, see fig.~\ref{fg2}.

For not too large number of atoms (see condition (\ref{deriv-cond}) below)
one can neglect the higher excited levels of the trap and approximate the
solution to the GP equation (\ref{eq1}) as follows
\begin{equation}
\Psi(x,\vec{r}_\perp,t) = f_u(x,t)\psi_u(\vec{r}_\perp) +
f_v(x,t)\psi_v(\vec{r}_\perp).
 \label{eq6}\end{equation}
This approximation allows for reduction of the GP equation to a system of
linearly coupled equations (equations (\ref{eq7}) below) for the
projections $f_{u,v}$ of the order parameter on the localized basis, due to
approximate decoupling of the basis functions in the nonlinear term. For
the double-well given in fig.~\ref{fg1}, for instance, we obtain the
following nonlinear cross-products:
\[
\frac{\int\mathrm{d}y\psi_u^2\psi_v^2}{\int\mathrm{d}y\psi_u^4} =
0.006, \quad
\frac{\int\mathrm{d}y\psi_u\psi_v^3}{\int\mathrm{d}y\psi_u^4} =
0.013, \quad
\frac{\int\mathrm{d}y\psi_u^3\psi_v}{\int\mathrm{d}y\psi_u^4} =
-0.040,\quad
\frac{\int\mathrm{d}y\psi_v^4}{\int\mathrm{d}y\psi_u^4} = 1.00.
\]

Substituting expansion (\ref{eq6}) into equation   (\ref{eq1}), multiplying
by the localized wave  functions, integrating over the transverse variables
$y$ and $z$, and throwing away small nonlinear cross-terms we arrive at the
coupled-mode system of Refs. \cite{tunnel1,tunnel2} but with the kinetic
terms:
\begin{subequations}
\label{eq7}
\begin{equation}
i\hbar \frac{\partial f_u}{\partial t} =
-\frac{\hbar^2}{2m}\frac{\partial^2 f_u }{\partial x^2} + ( E_u +
g_u|f_u|^2)f_u - K f_v, \label{eq7a}\end{equation}
\begin{equation}
i\hbar \frac{\partial f_v}{\partial t} =
-\frac{\hbar^2}{2m}\frac{\partial^2 f_v }{\partial x^2} + (E_v +
g_v|f_v|^2)f_v - K f_u. \label{eq7b}\end{equation}
\end{subequations}
 The zero-point energies $E_u$ and $E_v$ and the tunneling coefficient $K$
are defined as follows:
\begin{equation}
E_{u,v}
=\int\left(\frac{\hbar^2}{2m}(\nabla_{\vec{r}_\perp}\psi_{u,v})^2
+\mathcal{V}_\mathrm{ext}(\vec{r}_\perp)\psi_{u,v}^2\right)\mathrm{d}^2\vec{r}_\perp,\quad
g_{u,v} = \int g\psi_{u,v}^4\mathrm{d}^2\vec{r}_\perp,
\label{eq8}\end{equation}
\begin{equation}
K=-\int\left(\frac{\hbar^2}{2m}(\nabla_{\vec{r}_\perp}\psi_{u})(\nabla_{\vec{r}_\perp}\psi_v)
 + \psi_u
\mathcal{V}_\mathrm{ext}(\vec{r}_\perp)\psi_{v}\right)\mathrm{d}^2\vec{r}_\perp,
\label{eq9}\end{equation}
 here the $z$-dependence of the wave functions $\psi_u$ and
$\psi_v$ is given by the ground state wave function (the Gaussian) in the
$z$-direction (note that by changing the sign of either $\psi_u$ or
$\psi_v$, if necessary, one can always set $K>0$). Below, we assume that
the atomic interaction in the $v$-condensate is externally modified.

In the above analysis we have neglected the effect of nonlinearity on the
dynamics in the transverse ($y,z$)-plane assuming a small contribution from
the atomic interaction as compared to the energy gap $E_2 - E_1$ of the
trap. This leads to the following condition for each of the two parts of
the condensate:
\begin{equation}
\frac{|a_s| \mathcal{N}}{\ell} \ll 1,
 \label{deriv-cond}\end{equation}
where $\mathcal{N}$ is the number of atoms, $a_s$ is the scattering length,
and $\ell$ is the length of the condensate. Equation (\ref{deriv-cond}) is
the usual condition of applicability of the one-dimensional NLS equation
for the cigar-shaped condensate. The condition of applicability of the
classical mean-field approach is $|a_s|/\ell\ll1$ and $\mathcal{N}\gg1$
(see, for instance, Ref~\cite{QvsCc}), which is satisfied for large number
of atoms and condition (\ref{deriv-cond}). Finally, the general
applicability condition for our  two-mode approximation reads $K\ll
\hbar\omega_\perp$  (see the appendix).

In the calculations it is convenient to use a dimensionless form of system
(\ref{eq7}) with the following dimensionless variables:
\begin{equation}
\xi = \frac{x}{d_\perp},\quad d_\perp \equiv
\left(\frac{\hbar}{m\omega_\perp}\right)^{1/2}, \quad \tau =
\frac{\omega_\perp}{2}t, \quad \vare = \frac{2(E_v -
E_u)}{\hbar\omega_\perp},\quad \kappa =
\frac{2K}{\hbar\omega_\perp},\quad a = \frac{a_v}{|a_u|}.
 \label{eq10}\end{equation}
Here $a_u$ ($a_u<0$) and $a_v$ are defined as follows: $a_{u,v} \equiv
d^2_\perp\int a_s \psi_{u,v}^4 \mathrm{d}^2\vec{r}_\perp$  with $a_s$ being
the $s$-wave scattering length (different in the two condensates).
Introducing the dimensionless order parameters for the condensates in the
two wells,
\begin{equation}
u(\xi,\tau) = e^{i\frac{E_u t}{\hbar}}(8\pi|a_u|)^{1/2}f_u,\quad
v(\xi,\tau) = e^{i\frac{E_u t}{\hbar}}(8\pi|a_u|)^{1/2}f_v,
\label{eq11}\end{equation}
 we arrive at the coupled-mode system suitable for the numerical study:
\begin{subequations}
\label{eq12}
\begin{equation}
i \frac{\partial u}{\partial \tau} + \frac{\partial^2u}{\partial
\xi^2} + |u|^2u + \kappa v = 0, \label{eq12a}\end{equation}
\begin{equation}
i \frac{\partial v}{\partial \tau} + \frac{\partial^2v}{\partial
\xi^2} - (\vare  + a|v|^2)v + \kappa u = 0.
\label{eq12b}\end{equation}
\end{subequations}
The number of particles in the coupled-mode system (\ref{eq12}),  defined
as $N_u = \int\mathrm{d}\xi|u|^2$ and $N_v = \int\mathrm{d}\xi|v|^2$, is
related to the actual number of atoms $\mathcal{N}$  as follows:
$\mathcal{N}_{u,v} = (d_\perp/8\pi|a_u|) N_{u,v}$ (the coefficient
$d_\perp/8\pi|a_u|$ is of order $10^2-10^3$). Note that the quotient of the
number of atoms in the two condensates does not change under the scaling
transformation. Below we will use only the quantity $N_{u,v}$ referring to
it as ``the number of atoms'' for simplicity.

Finally, we note an important difference between our approach and that of
Ref.~\cite{couplmod}. In the latter work the nonlinear modes are used as
the basis functions, what results in an effective coupled-mode system with
nonlinear cross-terms, whereas our system (\ref{eq12}) consists of linearly
coupled equations.  Our coupled-mode system, however, applies only in the
case of weak tunneling through the barrier. But precisely in such setup the
experimental realization of control over the scattering length in a part of
the condensate may be feasible.


\section{Solitons in the tunnel-coupled condensates}
\label{sec3}

\subsection{General properties of the soliton solutions to the coupled-mode system}
\label{genpr}

 We are interested in the  stable stationary soliton solutions
to the coupled-mode system (\ref{eq12})  mainly for $a\ge0$, i.e. for a
repulsive-attractive pair of  condensates. Setting $u=e^{-i\mu\tau} U(\xi)$
and  $v=e^{-i\mu\tau} V(\xi)$,  where $\mu$ is the dimensionless chemical
potential and the functions $U(\xi)$ and $V(\xi)$ are real, we obtain the
corresponding stationary system for the soliton profiles:
\begin{subequations}
\label{eq15}
\begin{equation}
\mu U +  U_{\xi\xi} +  U^3 + \kappa  V = 0,
\label{eq15a}\end{equation}
\begin{equation}
(\mu - \vare)V +  V_{\xi\xi} - a V^3 + \kappa  U = 0.
\label{eq15b}\end{equation}
\end{subequations}
The two-mode approximation requires that $\kappa\ll1$ and
$|\varepsilon|\ll1$. On the other hand, system (\ref{eq15}) admits the
following scaling invariance:
\begin{equation}
 U = \sqrt{\kappa}\,\tilde{ U},\quad  V = \sqrt{\kappa}\,\tilde{
V},\quad \xi = \frac{\tilde{\xi}}{\sqrt{\kappa}},\quad \mu =
\kappa\tilde{\mu},\quad \vare = \kappa\tilde{\vare},
\label{eq16}\end{equation}
 where the new variables with tilde satisfy the same system with
$\tilde{\kappa} = 1$. For this reason we will keep $\varepsilon$ and
$\kappa$ arbitrary. The scale-invariance (\ref{eq16}) can be used to verify
the analytical solutions.

We are interested in the so-called fundamental solitons, i.e. when the
functions $U(\xi)$ and $V(\xi)$  are nodeless and decreasing exponentially
as $|\xi|\to\infty$. The exponential decay rate $k$, $U\sim e^{-k\xi}$ and
$V\sim e^{-k\xi}$ as $\xi\to\infty$,  is given by the dispersion law of the
linearized system. One obtains the following two branches:
\begin{equation}
k_{1,2} = \sqrt{\mu_{1,2} - \mu}, \quad \mu_1 = \frac{\vare}{2}
-\sqrt{\frac{\vare^2}{4} + \kappa^2},\quad \mu_2 = \frac{\vare}{2}
+\sqrt{\frac{\vare^2}{4} + \kappa^2}.
 \label{eq17}\end{equation}
Note that for arbitrary $\kappa$ and $\vare$ the boundary chemical
potentials satisfy the inequality $\mu_1<0<\mu_2$. They coincide with the
two quasi-degenerate  energy levels of the coupled-mode system (\ref{eq12})
when the nonlinearities are negligible.

For $a\ge0$ it can be shown that the two dispersion laws correspond to two
possible classes of the fundamental soliton solutions: $ U V>0$ or $ U V<0$,
below called the in-phase and out-phase solitons respectively. Indeed, the
amplitudes $ U_0\equiv U(0)$ and $ V_0\equiv V(0)$ are bounded as follows
\cite{PhysD}:
\begin{equation}
 U_0^2 < -2\mu,\quad \mathrm{if}\quad  U V>0;\qquad  a V_0^2 <
2(\mu -\vare),\quad \mathrm{if}\quad  U V<0.
\label{eq18}\end{equation}
 Noticing that the conditions  $k_1^2>0$ and $k_2^2>0$ demand
respectively $\mu<\mathrm{min}(0,\vare)$ and $\mu>\mathrm{max}(0,\vare)$ we
see that the in-phase solitons correspond to the $k_1$-branch, while the
out-phase ones pertain to the $k_2$-branch. Numerical calculations confirm
this prediction \cite{PhysD}.

We will discard the out-phase  soliton solutions from consideration, since
they bifurcate from zero at a higher chemical potential $\mu_2$
(\ref{eq17}). Therefore, it is clear that they can be stable only in
exceptional cases (they do not satisfy  stability criterion (\ref{eq24})
given below, consult also Ref.~\cite{A2}). Below by the soliton solution we
will mean the in-phase soliton.

When the condensates are attractive, i.e. for $a<0$, there is a class of
the sech-type soliton solutions \cite{PhysD}, which generalizes the
well-known symmetric solitons in the standard model of the dual-core
optical fiber \cite{A2,A1} (see also the review Ref.~\cite{ProgOpt}). The
sech-type soliton solution is given as
\begin{equation}
\left(\begin{array}{c} U\\  V\end{array}\right)
= \left(\begin{array}{c}A_{\mathrm{in}}\\
(-a)^{-1/2}A_{\mathrm{in}}\end{array}\right)\mathrm{sech}\left(
\frac{A_{\mathrm{in}}\xi}{\sqrt{2}}\right), \quad A_{\mathrm{in}}=
\sqrt{2\kappa}\left( -\mu/\kappa -(-a)^{-1/2}\right) ^{1/2}.
 \label{eq19}\end{equation}
It exists in the special case, when $\vare$, $\kappa$, $a$, and $\mu$  are
related as follows:
\begin{equation}
\vare = \kappa\left[ (-a)^{1/2}-(-a)^{-1/2} \right] \equiv
h(\kappa,a),\quad \mu < \mu_1 = -\kappa(-a)^{-1/2}.
 \label{eq20}\end{equation}
Evidently, the exact sech-type soliton solution does not exist for $a\ge0$.

There is another branch of the in-phase solitons, not known analytically,
which bifurcates from the sech-type soliton branch (\ref{eq19}), in the
domain of their existence, at the point
\begin{equation}
\mu_{\mathrm{bif}} = -\kappa\,\frac{4 - a}{3\sqrt{-a}},\quad \vare
= h(\kappa,a).
  \label{eq23}\end{equation}
For other parameter values the sech-type  solitons (\ref{eq19}) and the
bifurcating in-phase solitons coexist on separate curves $N = N(\mu)$
\cite{PhysD} (here $N$ is the total number of atoms in the condensate). At
least one of the two soliton branches must have continuation for positive
values of $a$, since we have found such in-phase solitons numerically in
Ref.~\cite{PhysD} (we have found exactly one branch for $a\ge0$).

It is important to know the stability properties of the soliton solutions.
In Ref.~\cite{PhysD} we have shown that for  $a\ge0$ the Vakhitov-Kolokolov
criterion for the soliton stability
\begin{equation}
\frac{\partial N}{\partial \mu}< 0
 \label{eq24}\end{equation}
can be applied (consult also Refs.~\cite{VK,Zakh}). The stability property
of the in-phase  solitons for $a<0$ is affected by bifurcations (i.e.
criterion (\ref{eq24}) does not apply for all $\mu$ when $a<0$). Details
can be found in Ref.~\cite{PhysD}.

\subsection{Analytical soliton solutions for $a=0$}
\label{sec3A}

For $a=0$ the self-bound states, solitons, are possible due to the balance
between the cubic nonlinearity in the attractive condensate and the linear
dispersion in both condensates. Thus we can use the following strategy to
find an approximate expression for the solitons. Equation (\ref{eq15b}),
which can  be cast as
\begin{equation}
\Lambda V \equiv \left(1 - \beta\frac{\mathrm{d^2}}{\mathrm{d}\xi^2}\right)
V = \kappa\beta U - a\beta V^3, \quad \beta \equiv (\vare - \mu)^{-1},
 \label{eq25}\end{equation}
can be explicitly resolved with respect to $V$ for $a=0$ using the method
of the Green function. It was already mentioned that for the in-phase
solitons $\mu<\vare$ and hence $\beta>0$. Thus  the operator $\Lambda$ is
positive definite  and invertible and the unique Green function reads
\mbox{$G(\xi) = \exp\{-|\xi|/\sqrt{\beta}\}/(2\sqrt{\beta})$.}  Therefore
the solution to equation (\ref{eq25}) for $a=0$ is
\begin{equation}
 V(\xi) = \frac{\kappa\sqrt{\beta}}{2}\int\limits_{-\infty}^\infty
e^{-\frac{|\xi-\eta|}{\sqrt{\beta}}} U(\eta)\mathrm{d}\eta.
 \label{eq27}\end{equation}
Solution (\ref{eq27}) can be expanded in the infinite series with respect
to the derivatives of $U(\xi)$:
\begin{equation}
 V(\xi) = \kappa\beta\left(  U(\xi) + \sum_{n=1}^\infty \beta^n
 U^{(2n)}(\xi)\right),
 \label{eq28}\end{equation}
which can be most easily verified by the formal inversion of the operator
$\Lambda$. The crucial step in the derivation of the approximate soliton
solution consists in retaining only two terms from the series (\ref{eq28})
when it is substituted into the equation for $U(\xi)$, equation
(\ref{eq15a}). For such a reduction  the following condition must be
satisfied
\begin{equation}
 U^3  \gg \kappa^2\beta^{n+1}
U^{(2n)},\quad n\ge2.
 \label{eq29}\end{equation}
Thus, under condition (\ref{eq29}), the coupled-mode system (\ref{eq15})
reduces to the canonical NLS
\begin{equation}
(\mu + \kappa^2\beta) U + (1+\kappa^2\beta^2) U_{\xi\xi} + U^3=0.
 \label{eq30}\end{equation}
In this approximation the $u$-component of the solution  is given by the
sech-type soliton:
\begin{equation}
 U = A\mathrm{sech}(k\xi),\quad A\equiv
\left[\frac{2(\mu_1-\mu)(\mu_2-\mu)}{\vare -
\mu}\right]^{1/2},\quad k \equiv
\left[\frac{(\vare-\mu)(\mu_1-\mu)(\mu_2-\mu)}{(\vare - \mu)^2
 +\kappa^2}\right]^{1/2}.
 \label{eq31}\end{equation}

Using the identities $\mu_1\mu_2 = -\kappa^2$ and $\vare = \mu_1 +\mu_2$ it
is easy to check that the amplitude $A$ satisfies the first inequality in
equation (\ref{eq18}) for all values of $\mu$. However, the analytical
decay rate $k$ is not the same as $k_1$ in equation (\ref{eq17}). This is
the price we pay for adopting the approximation.  In fact, as either
$\mu\to\mu_1$ or $|\mu|\gg|\mu_1|$ (recall that $\mu<\mu_1<0$) the decay
rate $k$ approaches $k_1$:
\begin{equation}
k = k_1\left[1 + \mathcal{O}(\mu_1-\mu)\right],\quad k =
k_1\left[1 + \mathcal{O}(\mu_1/\mu)\right].
 \label{eq32}\end{equation}
The applicability condition for the analytical approximation is then
satisfied at least in these two limits. Indeed, condition (\ref{eq29})
demands that $A^3 \gg \kappa^2\beta^3 k^4A$, where we have estimated
$U_\xi\sim k U$, and after a simple transformation we obtain
\begin{equation}
\alpha^2 \equiv  \frac{(\mu_1-\mu)(\mu_2-\mu)}{(\vare - \mu)^2 +
\kappa^2} \ll 1 + \frac{(\vare - \mu)^2}{\kappa^2}.
 \label{eq33}\end{equation}
Moreover, comparison with the numerical solution shows that the analytical
approximation captures all essential features of the number of atoms as
function of the chemical potential for {\it all} values of $\mu$ (see
fig.~\ref{fg3} below).

The $v$-component of the soliton, given by  equation (\ref{eq27}), cannot
be found  in the explicit form. Nevertheless, all the necessary physical
quantities, for instance, the number of atoms $N_v(\mu)$ and the energy of
the soliton, can be found in closed form. The shares $N_u$ and $N_v$ of the
BEC atoms trapped in the two wells read:
\begin{equation}
N_u = 4\left(\frac{(\mu_1-\mu)(\mu_2-\mu)[(\vare-\mu)^2+\kappa^2]}
{(\vare-\mu)^3}\right)^{1/2},
 \label{eq34}\end{equation}
\begin{equation}
N_v = 4\kappa^2\frac{(\mu_1-\mu)(\mu_2-\mu)}{(\vare-\mu)^{7/2}}
\sum_{n=0}^\infty\frac{(2n+1)\alpha +3}{[(2n+1)\alpha +1]^3},
 \label{eq35}\end{equation}
where $\alpha$ is defined in equation (\ref{eq33}).

The deformation of the curves $N_u(\mu)$ and $N_v(\mu)$ given by equations
(\ref{eq34}) and (\ref{eq35}) with variation of the zero-point energy
difference $\vare$ is illustrated in fig.~\ref{fg3} (we remind that the
scaled number of atoms $N = (8\pi|a_u|/d_\perp)\mathcal{N} $ is used). The
qualitative behavior of the analytical curves is in excellent agreement
with the numerical results of Ref.~\cite{PhysD} (the local maxima of the
analytical curves $N_{u,v}(\mu)$ are narrower and 10\% higher than the
maxima of the numerical curves).

\begin{figure}[ht]
\includegraphics{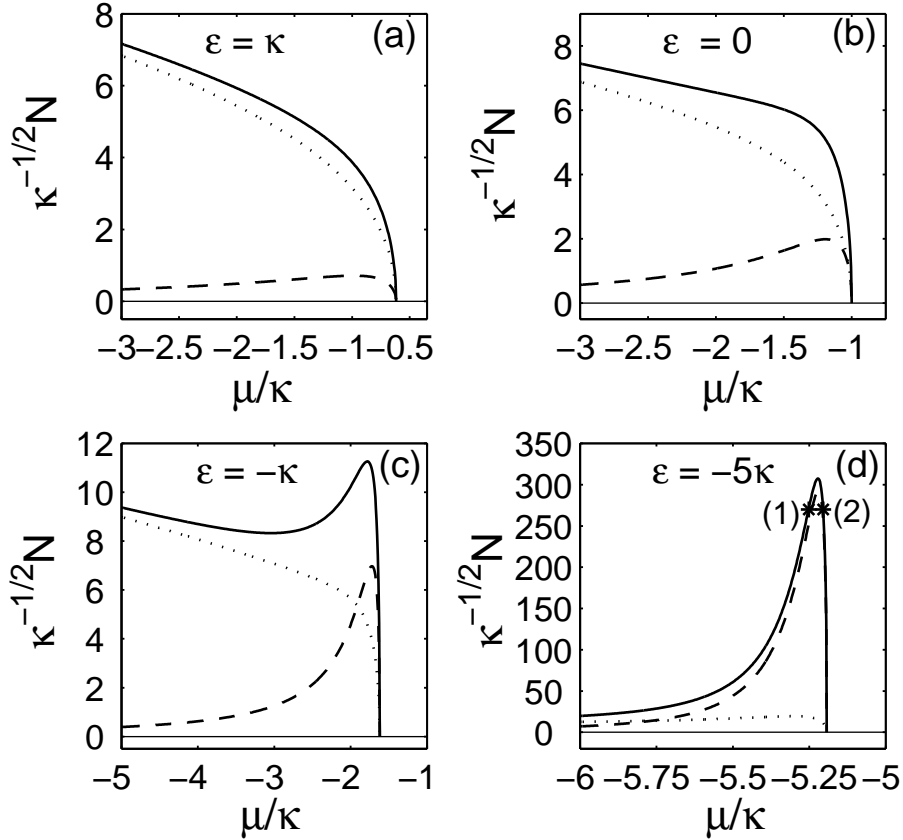}
\caption{\label{fg3} Behavior of the number of atoms in the two wells of
the trap  vs. the chemical potential with variation of the zero-point
energy difference between the wells (all quantities pertain to the
dimensionless system (\ref{eq12})). Here the dotted lines correspond to
$N_u$, the dashed lines to $N_v$, and the solid ones to the total number of
atoms $N = N_u+N_v$.  In panel (d) only the part of the picture about the
local maximum is shown (the full picture is similar to that in panel (c)).
}
\end{figure}

Note the striking sensitivity of the total number of atoms and that of the
fraction $N_v/N_u$ at the local maximum with respect to variation of the
zero-point energy difference $\vare$. From figs.~\ref{fg3}(c) and (d) it is
seen that a small variation of $\vare$ (with $\vare$ being on the order of
a small tunneling coefficient $\kappa$) results in a large variation of
both the total number of atoms and the relative share of atoms in the
$v$-condensate.

\begin{figure}[ht]
\includegraphics{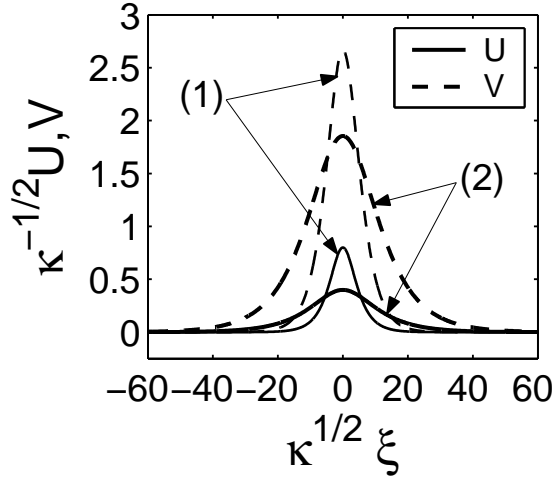}
\caption{\label{fg4} Unstable (1) and stable (2) soliton solutions to the
dimensionless system (\ref{eq12}) with the same total number of atoms which
correspond to the two stars in panel (d) of fig.~\ref{fg3}.}
\end{figure}

The solitons  corresponding to the two stars in fig.~\ref{fg3}(d) are shown
in fig.~\ref{fg4}. Soliton-(1) is unstable while soliton-(2) is stable. The
latter one represents the unusual self-bound state of the system, where
almost all atoms are gathered in the noninteracting condensate.

From fig.~\ref{fg3} it is seen that the soliton solution bifurcates from
the zero solution at $\mu = \mu_1$. In fact, for $a=0$, the total number of
atoms is indeed always equal to zero on the boundary of the soliton
existence. For $\mu\to\mu_1$ the sum on the r.h.s. of equation (\ref{eq35})
can be approximated as follows
\begin{equation}
\sum_{n=0}^\infty\frac{(2n+1)\alpha +3}{[(2n+1)\alpha +1]^3} =
\sqrt{\mu_2}(\mu_1-\mu)^{-1/2}+ \mathcal{O}\{(\mu_1-\mu)^{1/2}\},
\label{eq36}\end{equation}
 where have used the Euler-Maclaurin summation formula and
the definition of $\alpha$ (\ref{eq33}). Hence, the number of atoms is
proportional to $(\mu_1-\mu)^{1/2}$ as $\mu\to\mu_1$ and the bifurcation is
continuous for $a=0$  (in contrast, the number of atoms has a finite limit
as $\mu\to\mu_1$ for $a>0$ and $\vare = \vare_{\mathrm{cr}}$, see the next
section).

From equations (\ref{eq34})-(\ref{eq36}) one may deduce the ratio
of the slopes of  $N_u(\mu)$ and $N_v(\mu)$ as $\mu\to\mu_1$:
\begin{equation}
\lim_{\mu\to\mu_1}\frac{N_v}{N_u} = \frac{|\mu_1|}{\mu_2} =
\frac{\sqrt{(\vare/2)^2 +\kappa^2}-\vare/2}{\sqrt{(\vare/2)^2
+\kappa^2}+ \vare/2}.
 \label{eq37}\end{equation}
This is the quantitatively accurate result, since  in this limit condition
(\ref{eq33}) is well satisfied. Formula (\ref{eq37}) predicts, for
instance, that almost all atoms are gathered in the non-interacting
condensate for $|\,\vare|\gg\kappa$ and $\vare<0$.  Thus, to observe such
unusual solitons the zero-point energy difference $\vare$ should be
decreased sufficiently below zero. Physically, it is clear that the energy
decrease due to the attractive interaction in the $u$-condensate with
increase of $N_u$ should be compensated by a lower zero-point energy in the
$v$-well, if one wants to increase the ratio $N_v/N_u$.

Condition (\ref{eq33}) breaks down  about the local maximum of the curve $N
= N(\mu)$ (see fig.~\ref{fg3}(d)) and one has to use the whole infinite
series expansion (\ref{eq28}) for quantitatively correct description of the
soliton in the system. Therefore, the condensates are strongly coupled when
almost all atoms are trapped in the $v$-well.  We will see below that the
unusual solitons also appear for $0<a\ll 1$ and
$\vare_{\mathrm{cr}}<\vare<0$, with almost all atoms being contained in the
\textit{repulsive} condensate strongly coupled by the quantum tunneling to
the attractive one.

For $|\mu|\gg|\mu_1|$ we obtain the following estimates for the number of
atoms in the two wells:
\begin{equation}
N_u = 4\sqrt{-\mu} + \mathcal{O}(|\mu|^{-5/2}),\quad
\frac{N_v}{N_u} = 0.712 \frac{\kappa^2}{\mu^2} +
 \mathcal{O}\left(\mu^{-3}\right), \label{eq39}\end{equation}
where we have used  $\alpha = 1$ to evaluate the infinite sum in equation
(\ref{eq35}) (since $\alpha = 1 +\mathcal{O}(\mu^{-1})$).
 First of all,  almost all atoms are gathered in
the attractive $u$-condensate. Second, in this limit the $u$-component of
the soliton solution asymptotically satisfies equation (\ref{eq15a}) for
$\kappa = 0$, since the amplitude and decay rate of soliton (\ref{eq31})
asymptotically approach those for the soliton solution of a single NLS
equation: $A = \sqrt{-2\mu} + \mathcal{O}(|\mu|^{-3/2})$ and $k =
\sqrt{-\mu} + \mathcal{O}(|\mu|^{-3/2})$. Thus, for large negative $\mu$
the coupled-mode system predicts weak coupling between the two condensates.

To determine the  ground state of the system let us find the energy of the
soliton solutions. The GP equation (\ref{eq1}) corresponds to the energy
functional
\begin{equation}
\mathcal{E} = \int\mathrm{d}^3\vec{r}\left\{
\frac{\hbar^2}{2m}|\nabla\Psi|^2 +
\mathcal{V}_{\mathrm{ext}}|\Psi|^2 +\frac{g}{2}|\Psi|^4\right\}.
 \label{eq40}\end{equation}
In the coupled-mode approximation we have the following expression for the
energy
\begin{equation}
\mathcal{E} = E_u \mathcal{N} + \frac{\hbar\omega_\perp
d_\perp}{16\pi|a_u|} H , \label{eq41}\end{equation}
 where $\mathcal{N}$ is the actual total number of atoms and $ H $
is the Hamiltonian for system  (\ref{eq12}):
\begin{equation}
 H  = \int\mathrm{d}\xi\left\{ \left|\frac{\partial
u}{\partial\xi}\right|^2 + \left|\frac{\partial
v}{\partial\xi}\right|^2 +\vare|v|^2 - \kappa(uv^* + vu^*) -
\frac{|u|^4}{2} + \frac{a|v|^4}{2}\right\}.
 \label{eq42}\end{equation}
The Hamiltonian $H$ evaluated on the soliton solution reads
\begin{equation}
H_{\mathrm{sol}}  =
-\frac{4}{3}\left[\frac{(\mu_1-\mu)^3(\mu_2-\mu)^3}{(\vare-\mu)^2 +
\kappa^2}\right]^{1/2}\frac{(\vare-\mu)^2 + 2\kappa^2}{(\vare-\mu)^{5/2}} -
4\kappa^2\frac{(\mu_1-\mu)(\mu_2-\mu)(2\vare - 3\mu)}{(\vare -
\mu)^{7/2}}S(\mu)
 \label{eq43}\end{equation}
 where $S(\mu)=\sum_{n=0}^\infty \frac{(2n+1)\alpha +
\sigma}{[(2n+1)\alpha +1]^3}$ and $\sigma = \frac{2\vare - 5\mu}{2\vare -
3\mu}$.
 Taking into account that $2\vare > \mathrm{max}(3\mu,5\mu$), since $\mu<\mu_1$,
it is easy to see that $H_{\mathrm{sol}}$ is always negative. In the
coupled-mode approximation, expression (\ref{eq43}) is quantitatively
correct for $\mu\to\mu_1$ and $|\mu|\gg|\mu_1|$, and for other values of
$\mu$ captures the qualitative behavior of the energy.  Note that $ H $
scales as $\kappa^{3/2}$.

\begin{figure}[ht]
\includegraphics{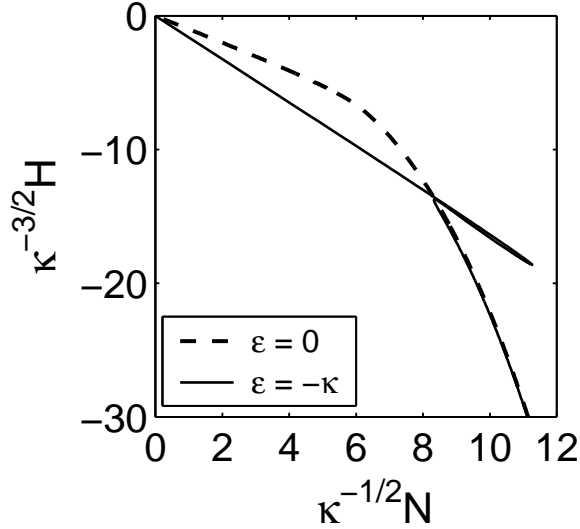}
\caption{\label{fg5} The energy vs. the total number of atoms (for the
dimensionless system (\ref{eq12})) for two values of the zero-point energy
difference: $\vare = 0$ (the dashed line) and $\vare =-\kappa$ (the solid
line).  The solid line has a very narrow twist without self-intersection,
which occupies the interval $8< N/\sqrt{\kappa} <12$. }
\end{figure}

Lowering $\vare$  below zero forces the energy to develop a twist, see
fig.~\ref{fg5}. The twisted part of the curve widens by decreasing $\vare$,
while its left endpoint moves towards $N=0$. (The curve $H = H(N)$ has the
shape similar to that of $\mu = \mu(N)$ after the following substitution is
applied to the latter: $(\mu_1-\mu)\to - H $.) It follows that, for almost
all values of $\vare$ and the total number of atoms,  the energy is
minimized on the solutions with large negative $\mu$ (in fig.~\ref{fg5} the
latter solitons correspond to the lower part of the solid line below the
twist, while the unusual solitons pertain  to the twisted part). However,
according to criterion (\ref{eq24}) the solitons with the larger share of
atoms in the $v$-condensate are stable (precisely, the solitons which
pertain to the upper part of the twist in fig.~\ref{fg5}), thus they can be
experimentally observed.

Finally, let us check condition (\ref{deriv-cond}). Estimating the
condensate length as $\ell/d_\perp\sim k$, we get that in the dimensionless
variables the one-dimensionality condition for the $u$-condensate is
$kN_u\ll 1$. Using the explicit expressions  (\ref{eq31}) and (\ref{eq34})
we arrive at $(\mu_1-\mu)(\mu_2-\mu) \ll \varepsilon -\mu$ which is
satisfied for $\mu$ close to $\mu_1$, i.e. in the domain of the unusual
solitons. Similar, for the $v$-condensate we have $k|a|N_v\ll 1$, which is
satisfied for small $a$ in the same range of $\mu$. On the other hand, the
soliton solutions with large negative $\mu$  may violate the
one-dimensionality condition (it is violated in the limit $|\mu|\gg
|\mu_1|$). It is a reflection of the fact that for large enough number of
atoms  the transverse degrees of freedom start to play a significant role
for the weakly coupled condensates, i.e. for the ground state in the
system, leading to the collapse instability at some $kN = f(k)$ (for
details on collapse consult Ref.~\cite{collapse}). However, for the number
of atoms below the collapse instability threshold the ground-state soliton
of the coupled-mode system is stable (but may have a modified shape), while
the unusual solitons are effectively one-dimensional, since they always
satisfy condition (\ref{deriv-cond}).

\subsection{Solitons for $a>0$ and the boundary of the bifurcation from zero}
\label{sec3B}

We have seen that for $a=0$ the soliton solutions bifurcate from the zero
solution at $\mu = \mu_1$ and the unusual stable solitons appear close to
the bifurcation boundary.  For $a>0$ the competition between the zero-point
energy difference $\vare$ and repulsion in the $v$-condensate creates a
lower boundary $\vare_{\mathrm{cr}} = b(k,a)$ for the bifurcation from
zero. Thus for $a>0$ the unusual solitons appear only for $\vare >
\vare_{\mathrm{cr}}$. Existence of such a boundary was discovered
numerically in Ref.~\cite{PhysD} for $a=1$.

It turns out that $\vare_{\mathrm{cr}}(k,a)$ can be found analytically in
the \textit{exact} form. To this goal we suppose that the solitons do
bifurcate from zero and find the value of $\vare$ for which we have a
contradiction. Setting $\epsilon = \mu_1-\mu$, we assume that $U =
\mathcal{O}(\epsilon^p)$ and $V =\mathcal{O}(\epsilon^q)$ as
$\epsilon\to0$, for some $p,q>0$. Using the estimate
${\mathrm{d}^2}/{\mathrm{d}\xi^2} \sim k^2_1 = \epsilon$ and  $\beta =
(\mu_2 + \epsilon)^{-1} = \mathcal{O}(1)$ in equation (\ref{eq25}) we
immediately obtain $q=p$. Note that the terms containing $ U$ in equation
(\ref{eq15a}) have the following powers of $\epsilon$: $\mu U =
\mathcal{O}(\epsilon^p)$, $U_{\xi\xi} = \mathcal{O}(\epsilon^{p+1})$, and
$U^3 = \mathcal{O}(\epsilon^{3p})$, where we have used the above estimate
for the second derivative. Therefore, irrespectively of the value of $p$,
it is clear that if an expression for $V$ contains all the terms with
powers of $\epsilon$  up to $\mathrm{max}(p+1,3p)$, it then contains all
the necessary terms we need to derive the soliton solution in the lowest
order of $\epsilon$.

The most general expression for $V$ that explicitly contains all the powers
of $\epsilon$ up to $\mathrm{max}(p+1,3p)$ is as follows
\begin{equation}
V = F( U) + C U_{\xi\xi} +\mathcal{O}(\epsilon^s),\quad
s\equiv\mathrm{min}(3p+1,p+2),
  \label{eq47}\end{equation}
where the function $F(U)$ and the constant $C$ (of order 1) are determined
by solution of equation (\ref{eq25}) in the required order of $\epsilon$.
We will need the expression for $V$ up to the terms of order
$\epsilon^{5p}$ only. We obtain
\begin{equation}
V = \kappa\beta U - a\kappa^3\beta^4 U^3 + 3a^2 \kappa^5\beta^7U^5 +
\kappa\beta^2 U_{\xi\xi} + \mathcal{O}(\epsilon^r),\quad r \equiv
\mathrm{min}(3p+1,p+2,7p).
 \label{eq50}\end{equation}
Here the coefficient $\beta = (\mu_2+\epsilon)^{-1}$ also should be
expanded into powers of $\epsilon$.

Substitution of expression (\ref{eq50}) into equation (\ref{eq15a}) leads
to the following equation for the $u$-component of the soliton:
\begin{equation}
-\epsilon\left(1 - \frac{\mu_1}{\mu_2}\right)U + \left(1 -
\frac{\mu_1}{\mu_2}\right)U_{\xi\xi} + \left(1 -
\frac{a\kappa^4}{\mu_2^4}\right)U^3 + \frac{3a^2\kappa^6}{\mu_2^7}U^5 =
\mathcal{O}(\epsilon^r).
 \label{eq51}\end{equation}

From equation (\ref{eq51}) it is quite clear that the bifurcation from zero
is forbidden if the coefficient at the cubic nonlinear term is negative (we
remind that $\mu_2>\mu_1$), i.e. if the lowest-order nonlinearity is
repulsive. In the latter case equation (\ref{eq51}) does not contain bright
soliton solutions in the limit $\epsilon\to0$.  Hence,  the exact boundary
of the bifurcation from zero is given as $\mu_2(\vare_{\mathrm{cr}},\kappa)
= \kappa a^{1/4}$. Using the definition of $\mu_2$ (\ref{eq17}), we obtain
the critical value of the zero-point energy difference
\begin{equation}
\vare_{\mathrm{cr}} = \kappa\left( a^{1/4} - a^{-1/4} \right).
 \label{eq52}\end{equation}
There are no in-phase solitons bifurcating from the zero solution below the
boundary $\vare_{\mathrm{cr}}$.

The properties of solitons close to the bifurcation boundary $\mu_1$  are
different in the two cases $\vare>\vare_{\mathrm{cr}}$ and
$\vare=\vare_{\mathrm{cr}}$. In the former case equation (\ref{eq51})
reduces to the canonical NLS equation (in this case $p=1/2$) and the
corresponding soliton solution reads
\begin{equation}
U = A_0\sqrt{\epsilon}\,\mathrm{sech}(\sqrt{\epsilon}\xi) +
\mathcal{O}(\epsilon^{3/2}),\quad A_0\equiv
\left[\frac{2\mu_2^3(\mu_2-\mu_1)}{\mu_2^4 -
a\kappa^4}\right]^{1/2}.
 \label{eq53}\end{equation}
The $v$-component of the soliton, obtained from equation (\ref{eq50}),  in
the same order in $\epsilon$ reads
\begin{equation}
V = \frac{\kappa}{\mu_2}U + \mathcal{O}(\epsilon^{3/2}) =
\frac{\kappa
A_0}{\mu_2}\sqrt{\epsilon}\,\mathrm{sech}(\sqrt{\epsilon}\xi) +
\mathcal{O}(\epsilon^{3/2}).
 \label{eq54}\end{equation}
It is straightforward to verify by integration that the number of atoms
corresponding to the soliton solution (\ref{eq53})-(\ref{eq54}) goes to
zero as $\epsilon\to0$:
\begin{equation}
N_u=\frac{4\mu_2^3(\mu_2 - \mu_1)}{\mu_2^4 - a\kappa^4}\sqrt{\epsilon} +
\mathcal{O}(\epsilon^{3/2}),\quad N_v=\frac{4\kappa^2\mu_2(\mu_2 -
\mu_1)}{\mu_2^4 - a\kappa^4}\sqrt{\epsilon} + \mathcal{O}(\epsilon^{3/2}).
 \label{eq55}\end{equation}
We conclude this case noticing that for $\vare>\vare_{\mathrm{cr}}$ the
solitons bifurcate from the zero
solution continuously.

It is interesting to note that if $a<0$ and $\vare = h(\kappa,a)$, i.e. as
in equation (\ref{eq20}), the soliton solution given by equations
(\ref{eq53})-(\ref{eq54}) coincides with the sech-type soliton solution
(\ref{eq19}) (in this case $\mu_1 = -\kappa (-a)^{-1/2}$ and  $\mu_2 =
\kappa (-a)^{1/2}$). Therefore, in this special case,  one can think of the
family of solitons (\ref{eq53})-(\ref{eq54}) as a continuation of the
sech-type soliton solutions (\ref{eq19}) for $a\ge0$.

In the critical case, $\vare = \vare_{\mathrm{cr}}$, the cubic term is
absent from equation (\ref{eq51}). The effective equation for the
$u$-component of the soliton is the quintic NLS equation (now $p=1/4$). The
corresponding soliton solution reads
\begin{equation}
U = B_0\epsilon^{1/4}\,\mathrm{sech}^{1/2}(2\sqrt{\epsilon}\xi) +
\mathcal{O}(\epsilon^{3/4}), \quad B_0 \equiv
\left[\frac{\kappa(1+a^{1/2})}{a^{3/4}}\right]^{1/4}.
 \label{eq56}\end{equation}
Here we have used the relations $\mu_2 = \kappa a^{1/4}$ and $\mu_1 =
-\kappa a^{-1/4}$ which follow from equation (\ref{eq17}) for $\vare =
\vare_{\mathrm{cr}}$. The $v$-component of the soliton is given by
\begin{equation}
V = \frac{\kappa}{\mu_2}U + \mathcal{O}(\epsilon^{3/4}) =
a^{-1/4}B_0\sqrt{\epsilon}\,\mathrm{sech}^{1/2}(2\sqrt{\epsilon}\xi)
+ \mathcal{O}(\epsilon^{3/4}).
 \label{eq57}\end{equation}
 Though soliton solution (\ref{eq56})-(\ref{eq57})  bifurcates
from the zero solution, since both components tend to zero in the limit
$\epsilon\to0$, the corresponding number of atoms now approaches a
\textit{finite} limiting value. Indeed, we obtain:
\begin{equation}
N_u=\frac{\pi}{2}\left[\frac{\kappa(1+a^{1/2})}{a^{3/4}}\right]^{1/2}
+ \mathcal{O}(\epsilon^{3/2}),\quad
N_v=\frac{\pi}{2}\left[\frac{\kappa(1+a^{1/2})}{a^{7/4}}\right]^{1/2}
+ \mathcal{O}(\epsilon^{3/2}).
 \label{eq58}\end{equation}
The critical bifurcation can thus be rightfully called a discontinuous one,
since the total number of atoms must be greater than some threshold  value
for the system to have a soliton solution of vanishing amplitude.

\begin{figure}[ht]
\includegraphics{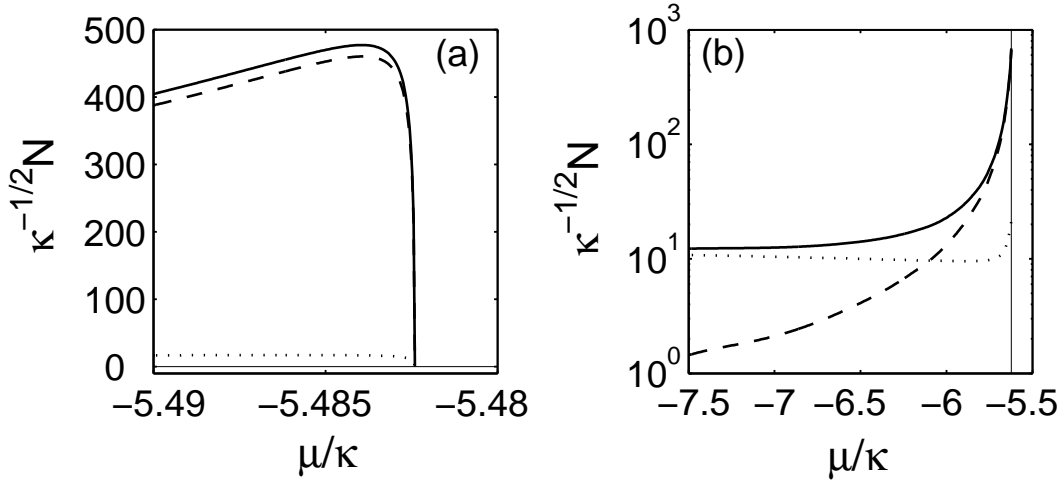}
\caption{\label{fg6}  Bifurcation of the soliton solutions for two values
of the zero-point energy difference: $\vare = -5.30\kappa$ (panel (a)) and
$\vare_{\mathrm{cr}} = -5.45\kappa$ (panel (b)). The dimensionless units of
system (\ref{eq12}) are used. Here $a=0.001$. In both panels: the dotted
lines correspond to $N_u$, the dashed to $N_v$, and the solid ones to the
total number of atoms. }
\end{figure}

In fig.~\ref{fg6} we show the numerically computed behavior of the number
of atoms vs. chemical potential for $a=0.001$ in the two cases:
$\vare>\vare_{\mathrm{cr}}$ (panel (a), $\vare = -5.3\kappa$) and on the
boundary of the bifurcation from zero $\vare=\vare_{\mathrm{cr}}$ (panel
(b), $\vare_{\mathrm{cr}} = -5.45\kappa$).

It is seen that the curves shown in fig.~\ref{fg6}(b) indeed  cease at the
finite values $N_u = 21$ and $N_v = 670$ (according to equation
(\ref{eq58})) on the boundary $\mu=\mu_1$ of the soliton existence.  The
corresponding soliton solutions are unstable  as $\mu\to\mu_1$, as they
correspond to the positive slope of the curve $N = N(\mu)$ for the total
number of atoms. Solitons corresponding to fig.~\ref{fg6}(a), on the
contrary, are stable in this limit.

Moreover, for $\varepsilon <\varepsilon_\mathrm{cr}$,  for values of the
chemical potential close to the boundary $\mu_1$  there are soliton
solutions which represent the unusual self-bound states of BEC, where
almost all atoms are gathered in the repulsive $v$-condensate.  The unusual
solitons have the shape similar to that of soliton-(2) of fig.~\ref{fg4}
and appear for small values of the parameter $a$.  Indeed, their appearance
is conditioned by $\vare < 0$ at least, according to formula (\ref{eq37}),
while the condition  $a<1$ is necessary for a negative lower boundary
$\vare_{\mathrm{cr}}$. Obviously, the stability of such solitons is derived
from the  strong tunnel-coupling to the attractive $u$-condensate and the
atoms can be thought as existing in the ``superposition'' of the repulsive
and attractive pairs.

\section{Conclusion}
\label{sec4}

We have studied the solitons in BEC in  a double-well trap with
far-separated wells, when the atomic interaction is externally switched
from attractive to repulsive in the condensate trapped in one of the wells.

First of all, there are soliton solutions with almost all atoms contained
in  the attractive condensate, which correspond to  weak tunnel-coupling
between the condensates in the two wells of the trap. In this case the
attractive condensate serves as a spatially inhomogeneous  source of atoms
for the repulsive one, thus tailoring the repulsive condensate to the
soliton form.

However, there are unusual stable solitons in the system, when almost all
atoms are contained in the repulsive condensate. It is found that such
soliton solutions appear for weak repulsion when the zero-point energy of
the well with the repulsive condensate is lower than that of the attractive
one but is above some critical value depending on the tunneling and
interaction coefficients. The unusual  solitons describe the condensates
strongly coupled via the quantum tunneling, notwithstanding the fact that
the tunneling coefficient is small. This apparent paradox is resolved by
noticing that the tunneling probability is proportional to the number of
atoms and that the total number of atoms increases with increase of the
relative share of atoms in the repulsive condensate.

The ground state of the system corresponds to the soliton solution with the
larger share of atoms contained in the attractive condensate. However, the
solitons with the larger share of atoms in the repulsive condensate are
stable, hence they  can be experimentally observed.

\section*{Acknowledgements}
This work was supported by the FAPESP, FAPEAL and the CNPq grants.
V.S. would like to thank the Departamento de F\'{\i}sica,
Universidade Federal de Alagoas, in Macei\'o for kind hospitality.

\appendix*
\section{Wave functions with the localization property }

Here we  show that the inverse-proportionality relation (\ref{eq4}) holds
for the eigenfunctions corresponding to the quasi-degenerate energy levels
in the double-well trap with weak tunneling through the barrier.

The weakly asymmetric double-well potential can be cast as $\mathcal{V} =
\mathcal{V}_S(y) + \mathcal{V}_A(y)$, where $\mathcal{V}_S(-y)=
\mathcal{V}_S(y)$ is the symmetric potential and $\mathcal{V}_A(-y) =
-\mathcal{V}_A(y)$ is the perturbation. Under assumption (\ref{conda1}) the
eigenfunctions of the full Hamiltonian
\begin{equation}
\hat{H} = \hat{H}_S + \mathcal{V}_A(y) \equiv
-\frac{\hbar^2}{2m}\frac{\mathrm{d}^2}{\mathrm{d}y^2} +
\mathcal{V}_S(y) +\mathcal{V}_A(y),
 \label{eqa1}\end{equation}
corresponding to the shifted degenerate energy levels are given by a
rotation of the wave functions $|0\rangle$ and $|1\rangle$ corresponding to
the degenerate energy levels of the unperturbed Hamiltonian $\hat{H}_S$:
\begin{equation}
|\mathcal{E}_0\rangle = \frac{|0\rangle + \mathcal{A}|1\rangle}{\sqrt{1 +
\mathcal{A}^2}},\quad |\mathcal{E}_1\rangle = \frac{\mathcal{A}|0\rangle -
|1\rangle}{\sqrt{1 + \mathcal{A}^2}}, \quad \mathcal{A} \equiv
\frac{\Delta\mathcal{E} - \Delta E}{2K},
 \label{eqa4}\end{equation}
where $\Delta E = E_1-E_0$ and $\Delta \mathcal{E} =
\mathcal{E}_1-\mathcal{E}_0$ is the energy spitting in the unperturbed and
perturbed trap, respectively, and $K = \langle 0|\mathcal{V}_A|1\rangle$.
The condition for equation (\ref{eqa4}) is $K\ll \hbar\omega_\perp$, where
$\hbar\omega_\perp$ serves as an estimate on the energy gap $E_2-E_1$ in
the double-well trap.

Equation (\ref{eqa4}) guarantees  that if the inverse proportionality
relation (\ref{eq4}) holds for the symmetric double-well it also holds for
the asymmetric one. This follows from a simple calculation:
\begin{equation}
\frac{\langle y_-|\mathcal{E}_0\rangle}{\langle y_-|\mathcal{E}_1\rangle} =
 \frac{\frac{\langle y_-|0\rangle}{\langle y_-|1\rangle} +
\mathcal{A}}{\mathcal{A}\frac{\langle y_-|0\rangle}{\langle y_-|1\rangle} -
1}= \frac{-\frac{\langle y_+|1\rangle}{\langle y_+|0\rangle}
+\mathcal{A}}{-\mathcal{A}\frac{\langle y_+|1\rangle}{\langle y_+|0\rangle}
- 1}= - \frac{\langle y_+|\mathcal{E}_1\rangle}{\langle
y_+|\mathcal{E}_0\rangle},
 \label{eqa8}\end{equation}
 where the supposed inverse proportionality relation
(\ref{eq4}) for $|0\rangle$ and $|1\rangle$ have been used. On the other
hand, under the assumption that the tunneling coefficient is weak the
eigenfunctions of the symmetric double well are well-known combinations of
the eigenfunctions for the two far-separated wells: $|0\rangle =
(|\ell\rangle + |r\rangle)/\sqrt{2}$, $|1\rangle = (|\ell\rangle
-|r\rangle)/\sqrt{2}$. Then the relation $\frac{\langle
y_-|0\rangle}{\langle y_-|1\rangle} \approx -\frac{\langle
y_+|1\rangle}{\langle y_+|0\rangle} \approx 1$ is obvious due to
localization of the wave functions $|\ell\rangle$ and $|r\rangle$ in the
respective wells and the symmetry. This concludes the proof of equation
(\ref{eq4}).

\end{document}